\documentclass[12pt,preprint]{aastex}
\usepackage{graphicx}
\begin{document}

%%-----------------------------
%%      the top matter
%%-----------------------------
\title{Accretion--ejection phenomena from young stars}
\author{Jonathan Ferreira}\affil{Laboratoire d'Astrophysique de
Grenoble, 414 rue de piscine, 38041 Grenoble, France \\
Jonathan.Ferreira@obs.ujf-grenoble.fr}
\begin{abstract}
    In the current paradigm of star formation magnetic fields play
a very central role. Indeed, they probably help or even channel the
initial gravitational collapse of the parent molecular cloud.  But
their most spectacular effect is certainly the production of
ubiquitous, powerful, bipolar self-collimated jets.  Also, in the last
decade, evidence has come that in such systems a magnetosphere links
the protostar to its surrounding accretion disc.

I will first briefly review the magnetohydrodynamic models of jets from young
stars.  Then, I will discuss some constraints on the magnetospheric
interaction since accretion must proceed despite strong stellar magnetic
fields.  Finally, I will introduce the 3Reconnection X-wind" scenario
that leads to episodic jet formation with efficient removal of
protostellar angular momentum.
\end{abstract}
%
%\maketitle
%%-----------------------------
%% your text: 10 pages max
%%-----------------------------
 % 

%%%%%%%%%%%%%%%%%%%%%%
\section{Introduction}

Collimated ejection of matter is widely observed in several astrophysical
objects: inside our own galaxy from all young forming stars (Hartigan et al.
\cite{hart95}) and some X-ray binaries
(Mirabel and Rodriguez \cite{mir99}), but also from the core of active
galaxies (Cao and Jiang \cite{cj99}, Jones et al. \cite{jones00}).  All
these objects share the following properties: jets are almost
cylindrical in shape; the presence of jets is correlated with an
underlying accretion disc surrounding the central mass; the total jet
power is a sizeable fraction of the accretion power.

Most of observed images show jets that are extremely well collimated
{\em already  close to the source}, with an opening angle of only some
degrees.  On the other hand, the derived physical conditions show that
jets are highly supersonic.  Indeed, emission lines require a
temperature of order $10^4$ K, hence a sound speed $c_s \sim 10$
km/s while the typical jet velocity is $v_j\sim 300$ km/s.  The
opening angle $\theta$ of a ballistic hydrodynamic flow being simply
$\tan \theta = c_s/v_j$, this provides $\theta\sim 5^o$ for YSOs,
nicely compatible with observations.  Thus, observed jets could well
be ballistic.  Note that this qualitative argument is not changed by
recent observations of rotation in jets, since the rotation velocity
is of the same order than the sound speed (Testi et al. \cite{testi02}).

Therefore, the fundamental question is {\em how does a physical system
produce an unidirectional supersonic flow~?} This implies that
confinement must be intimately related to the acceleration process. 
To date, the only physical process proved to be capable of accelerating plasma
along with a self-confinement relies on the action of a large scale
magnetic field (carried along by the jet).  Such a magnetic field is assumed to 
arise from either advection of interstellar (fossil) magnetic field or from
local dynamo (or both).

There are three different situations potentially capable,
at all evolutionary stages\footnote{The evolutionary scenario of a
young forming star goes from Classes 0, I (embedded source) to Class
II (optically visible Classical T~Tauri star) and Class III (Weak line
T~Tauri star: no signs of accretion/ejection).  The correlation
between accretion and ejection remains valid from Class 0 to Class II
phases (Cabrit et al. \cite{cabrit92}, Hartigan et al. \cite{hart95},
Bontemps et al. \cite{bontemps96}).}, of driving magnetized jets from
young forming stars:\\
$\bullet$ {\bf The protostar alone}: stellar winds extract both energy and
mass from the protostar itself (Sauty et al. \cite{sauty02}). \\ 
$\bullet$ {\bf The accretion disc alone}: ``disc winds'' are produced from a
large radial extension in the disc and are fed with both mass and
energy provided by accretion (Casse and Ferreira \cite{cf2000a}).\\ 
$\bullet$ {\bf The interaction zone between the disc and the protostar}:
``X-winds'' are disc-winds produced in a tiny region around the
magnetopause between the disc and the protostar (Shu et al.
\cite{shu94}, Lovelace et al. \cite{love99}).

Pure stellar wind models are less favoured because observed jets carry
far too much momentum.  In order to produce an observed jet, a protostar should
be either much more luminous or rotating faster than observations show
(DeCampli \cite{dC81}, Konigl \cite{konigl86}).  Unless otherwise proven,
this leaves us with either disc-winds or X-winds.  Note that what
distinguishes these last two scenarii is mainly belief: belief that no
significant large scale magnetic field could be found in the innermost
disc region (up to a few au).  It is however possible to
observationaly discriminate between them (Shang et al.
\cite{shang02}, Garcia et al. \cite{garcia01b}).  Nevertheless, the
disc-wind scenario is usually favoured because it offers an
"universal" paradigm able to explain jets from several astrophysical
objects without relying on the central object.

%%%%%%%%%%%%%%%%%%%%%%%
\section{Accretion--ejection from circumstellar accretion discs}

\subsection{The disc ejection efficiency}

A large scale (mean) magnetic field of bipolar topology is assumed to
thread an accretion disc, allowing ejected plasma to flow along open field
lines. This field extracts both angular momentum and energy from the
underlying disc and transfers them back to a fraction of ejected plasma.  
This plasma is accelerated from the disc surface by the so-called
"magneto-centrifugal force" and, farther away, is self-collimated by
the magnetic "hoop-stress" (see Ferreira (\cite{f97}) for an explanation).
Accretion and ejection are therefore interdependent, which requires a new
theory of accretion discs.  Indeed, one must solve
the disc vertical structure\footnote{Jet models of eg. Blandford and
Payne \cite{bp82} or Pelletier and Pudritz \cite{pp92} treated the
disc as a boundary condition.} as well as the radial one, ie.  the
full 2D problem.  This is the reason why no toy-model has been able
yet to catch the main features of these accretion--ejection structures.

In these accretion discs, because mass is being lost in the jets, the
accretion rate varies with the axial distance such that 
\begin{equation}
   \dot M_a \propto r^{\xi}
\end{equation}
where $\xi$ measures the ejection efficiency (Ferreira and
Pelletier \cite{fp93}).  While the standard accretion disc model is
characterized by $\xi=0$, a revisited theory of ejecting discs
provides the allowed values of $\xi$ as a function of the disc
properties (Casse and Ferreira \cite{cf2000a}).  In turn, $\xi$ fixes the
amount of mass that is actually ejected by the accretion disc.  It
turns out that the overall jet behaviour (asymptotic jet radius and
velocity, degree of collimation) drastically depends on this load
(Ferreira \cite{f97}, Ouyed and Pudritz \cite{ouyed99}).  There has 
been quite a lot of work done on the area of jet launching from
keplerian accretion discs, but only very few addressed explicitely the
problem of mass loading into the jet\footnote{This holds also for
X-wind models.}, ie.  the value of $\xi$.  To my knowledge, apart from
the work reported here, only Wardle and Konigl (\cite{wk93}) and Li
(\cite{li95}) did it, but they used crude approximations forbidding
them to get the physically acceptable range in $\xi$ (see Ferreira
(\cite{f97}) for a detailled discussion).

\subsection{Self-similar models}

Answering this can only be done by constructing a self-consistent
accre\-tion--ejection model, from the resistive MHD accretion disc to the
ideal MHD jet. Dealing with the partial derivatives involves the
use of self-similar solutions\footnote{Self-similarity is a special
case of the method of separation of variables commonly used in
mechanics.  It allows to solve the full set of MHD equations without
any approximation.} following the scaling imposed by the gravity of
the central object (Ferreira and Pelletier \cite{fp93}).  The validity of such
solutions is questionable if jets arise from a small region in the
disc.  However, if jets are launched from a large region (say between
$0.1$ and a few au), then they provide a correct description.  In what
follows, I describe some key features of such accretion disc models
and recommend the interested reader to refer to the series of
published papers.

The accretion disc must be resistive enough so that matter, which is
both rotating and accreting, can indeed cross the magnetic field
lines.  Such a resistivity has to be anomalous and is expected to
arise from MHD turbulence.  Actually, the main assumption of the model
is that such a turbulence can indeed be described by local
phenomenological transport coefficients (resistivity, viscosity and
heat conductivity). 
Within this framework, it has been found that steady-state jets
require a magnetic field close to equipartition
($\frac{B^{2}}{\mu_{0}}\sim P$, where $P$ is the plasma pressure).  The
magnetic field cannot be stronger otherwise it would forbid ejection. 
Indeed, the vertical component of the Lorentz force pinches the disc
and the {\em only} force pushing matter up is the vertical gradient of
the plasma pressure.  This occurs at the disc surface where matter can
still cross the field lines (Ferreira and Pelletier \cite{fp95}).

For adiabatic or isothermal magnetic surfaces, the ejection efficiency
is always very small, typically $\xi \sim 0.01$ (Ferreira
\cite{f97}, Casse and Ferreira \cite{cf2000a}).  However, if some
additional heating occurs at the disc surface, enhancing there the
plasma pressure gradient, then much higher ejection efficiencies can
be reached, up to $\xi \sim 0.5$ (Casse and Ferreira \cite{cf2000b})~! 
This fact introduces a tremendous complexity in the theory, since
knowing the ejection efficiency $\xi$ requires now to solve a
realistic energy equation.  This is however very promising since it
offers a quite natural explanation of why different classes of
astronomical objects may produce jets with different efficiencies.

The theory of accretion discs driving jets is now well established, in
the sense that both the relevant physical processes and parameter
space are known.  However, there is still some "freedom" since the
ejection efficiency $\xi$ is actually determined by the unknown MHD turbulence
parameters (Casse and Ferreira \cite{cf2000a}, \cite{cf2000b}). 
Either one gets these parameters from a theory of MHD turbulence
inside discs (a dreadfull task), or one uses observations to infer
these values. 

This is the approach used by Garcia et al.  (\cite{garcia01a},
\cite{garcia01b}).  The energy equation and ionization equilibrium have
been computed along each magnetic field lines, using a self-similar
disc-wind model heated by ambipolar diffusion (neutrals-ions
collisions).  It was then possible to reproduce synthetic
observations: spatially resolved forbidden line emission maps,
long-slit spectra, as well as line ratios.  Line profiles and jet
widths appeared to be good tracers of the wind dynamics and
collimation, whereas line ratios essentially trace gas excitation
conditions.  All the above diagnostics were
confronted\footnote{Convolution by the observing beam is essential for
a meaningful test of the models.} to observations of T~Tauri star
microjets, with a very nice general agreement.  However, it was found
that jets with $\xi \sim 0.01$ are too tenuous and with too large
velocities\footnote{The maximum asymptotic velocity is $v_{j}\simeq
\Omega_{o}r_{o} \xi^{-1/2}$, where $\Omega_{o}r_{o}$ is the keplerian
speed at the footpoint of the magnetic field line.}.  Thus, a
detailled comparison with observations shows that T~Tauri stars jets
require higher disc ejection efficiencies, thereby discs with a warm
chromosphere or corona.  Such an effect may well be the natural
outcome of disc illumination by both UV and X-rays produced by the
accretion shock and star itself (see Feigelson and Montmerle \cite{feigel99}).

But one should also realize that "cold" jets, ie.  jets produced by the sole
magneto-centrifugal force, are an extravagant theoretical
simplification.  Indeed, there is no physical reason why thermal
effects should play no role, especially in view of its enormous
importance in the solar wind.  Moreover, at the surface of a
magnetized accretion disc, one expects strong energy dissipation due
to the presence of both small scale magnetic loops and differential
rotation.  Hence, any disc producing jets should have a magnetically
heated corona (see also Kwan \cite{kwan97}).

%%%%%%%%%%%%%%%%%%%%%%%%
\section{Accretion--ejection from the star--disc interface}

Even if accretion discs produce jets, only 1 to $\sim 10 \%$ of their
mass is being actually ejected.  Thus, most of the mass accreting through
the disc will eventually fall into the star.  Not much freedom is
therefore left for our imagination:\\
$\bullet$ {\bf If stellar magnetic fields are "weak"}, then we expect the
formation of an equatorial boundary layer between the disc and the star
(Regev and Bertout \cite{regev95}, Popham et al. \cite{popham96}).\\
$\bullet$ {\bf If stellar magnetic fields are "strong"}, then we expect the
disc plasma to be forced to follow the stellar magnetospheric field,
giving rise to so-called accretion curtains (or funnel flows, Edwards
et al. \cite{edwards94}).

\subsection{Why is a magnetic interaction zone necessary~?}

The "weak field" scenario has a drawback which may be quite severe. 
Indeed, all T~Tauri stars are observed to rotate at about $10 \%$ of
their break-up velocity (Bertout \cite{bertout89}).  This means that
matter reaching the equator is rotating much faster than the star. 
Thus, such a meridional boundary layer can only lead to a stellar
spin-up. But the current interpretation of
observations is that accreting T~Tauri stars (ie.  CTTS) maintain a
constant period and, once their disc has disappeared (ie.  WTTS),
their own gravitational contraction makes them spinning up (Bouvier
et al. \cite{coyotes97}, Bouvier, Forestini and Allain
\cite{bouvier97}).  This implies that the coupling between the star
and the disc must brake down the contracting star and remove both
accreting material and its own angular momentum.

One could argue that the star, viscously linked to the surrounding disc,
is simultaneously driving a wind that carries away this angular
momentum.  But to my knowledge, no conventional stellar wind model
ever showed that it could indeed produce such a strong magnetic
braking.  Moreover, if the star is able to produce such a stellar
wind, then it surely has also a magnetospheric field threading the
circumstellar disc.  One has therefore to explain why such a
configuration has no dynamical influence on the disc (no relevant
torque, no "freezing" action on the required MHD turbulence).

We are therefore lead to assume there must be a magnetospheric
interface between the star and its accretion disc.  Such a conclusion
is fortunately supported by several observational pieces of evidences:
measures of strong stellar magnetic fields (of order kG, Guenther
et al. \cite{guenther99}), emission lines probing infalling matter
from high latitudes (Beristain et al. \cite{beristain01}), photometric
variability and obscuration (Bouvier et al. \cite{bouvier99}), holes
in the innermost disc region (Muzerolle et al. \cite{muzerolle98} and
references therein).

\subsection{An "extended" or a "narrow" interface~?}

We saw that the disc and the star are dynamically linked by large scale
magnetic fields in such a way that the contracting star is being spun
down.  The question now is to find the correct magnetic configuration. 
Let us first introduce two important definitions:

$\bullet$ {\bf The corotation radius}, namely $r_{co}=
(GM_{*}/\Omega_{*}^{2})^{1/3}$, is defined as the radius where the
keplerian rotation is equal to the stellar one (note that disc
material is always slightly sub-keplerian).

$\bullet$ {\bf The magnetopause radius} $r_{m}$ is defined as the equatorial
radius below which there is no disc plasma anymore, only the force-free stellar
magnetosphere.

The respective location of these two radii plays an important role in
the star--disc interface dynamics.  There are two extreme magnetic
configurations that must allow accretion towards the central star and
provide a magnetic braking: (1) an "extended" interface, which is a situation
where stellar field lines thread the accretion disc on a wide range of
radii, from $r_{m}$ to a much larger outer radius; (2) a "narrow"
interface where this outer radius is (larger but) of the same order
than $r_{m}$.

The extended configuration assumes a (turbulent) disc magnetic
diffusivity so that stellar field lines can indeed thread the disc
over a large extension (Gosh and Lamb \cite{gl78}).  All stellar field
lines anchored in the disc at a radius smaller than $r_{co}$ produce a
spin-up, whereas those anchored beyond $r_{co}$ produce a spin-down. 
Inside this framework, magnetic braking of the star can in principle
be achieved if the magnetospheric star--disc link is such that the two
torques almost balance each other (Cameron and Campbell \cite{cc93}, Li
\cite{liJ96}).  Usually, it is found that $r_{m} < r_{co}$ but close
to it, otherwise the spin-up torque would be too large.  However, as Bardou and
Heyvaerts (\cite{bardou96}) showed, accretion inside the disc requires
a viscous torque greater than that due to the stellar field lines.  As
a consequence, the consistent magnetic diffusivity must be so large that,
instead of being maintained at the same radius (producing thereby a
spin-down torque), the magnetospheric field lines inflate (Aly's
theorem \cite{aly84}) and diffuse rapidly outside the disc.  The
configuration envisionned cannot simply be sustained.

It seems therefore more natural to expect a sharp transition from the
force-free magnetosphere to the star-disconnected accretion disc, ie.  a
"narrow" star--disc interface.  This is much harder to model since it
involves strong gradients in both radial and vertical directions. 
This sharp transition requires $r_{m} \geq r_{co}$ in order to provide
only a negative torque (magnetic braking), ie.  a disc truncated at
quite a large distance from the stellar surface $R_{\star}$ (the
minimum distance would be $r_{m} \simeq r_{co} \simeq 3 R_{\star}$). 
This arises only from the constraint that the star--disc interaction
must provide a magnetic braking {\em on stellar timescales} ($10^{5}$
to several $10^{6}$ yrs), much longer than the local disc dynamical
timescale (days).  How does the system maintain $r_{m} \geq r_{co}$ on
these long timescales is still an open question.

%%%%%%%%%%%%%%%%%%%%%%%
\section{The Reconnection X-wind model}

\subsection{Formation of a magnetic neutral line}

Both theoretical and observational arguments point towards the
following picture: accretion discs are truncated at $r_{m}\geq r_{co}$
by stellar magnetospheres whose radial extension inside the discs are
quite narrow.  There is however something weird about this picture. 
If the disc has to be the reservoir of the stellar angular momentum,
then how can accretion take place~?  Discs are much less massive than
the central star so it is quite hard to imagine they will cope with
the huge stellar torque and quietly carry away (by radial viscous
transport) the excess angular momentum deposited by the star.  In the
X-wind model, such a transport is assumed to be provided by viscosity 
around the magnetopause, so that this excess is finally carried away
out of the disc by the X-wind (Shu et al. \cite{shu94}).  But this X-wind
is nothing more than a disc wind, capable only to transport the exact
amount of angular momentum that allows disc plasma to accrete.  In
other words, {\em one cannot drive a jet with stellar rotational energy if
the field lines are anchored in the disc}.  So, understanding the
accretion--ejection physical process leads us to favour a "stellar wind" as the
final reservoir for the stellar angular momentum.

\begin{figure} 
    \centerline{\includegraphics[height=5cm]{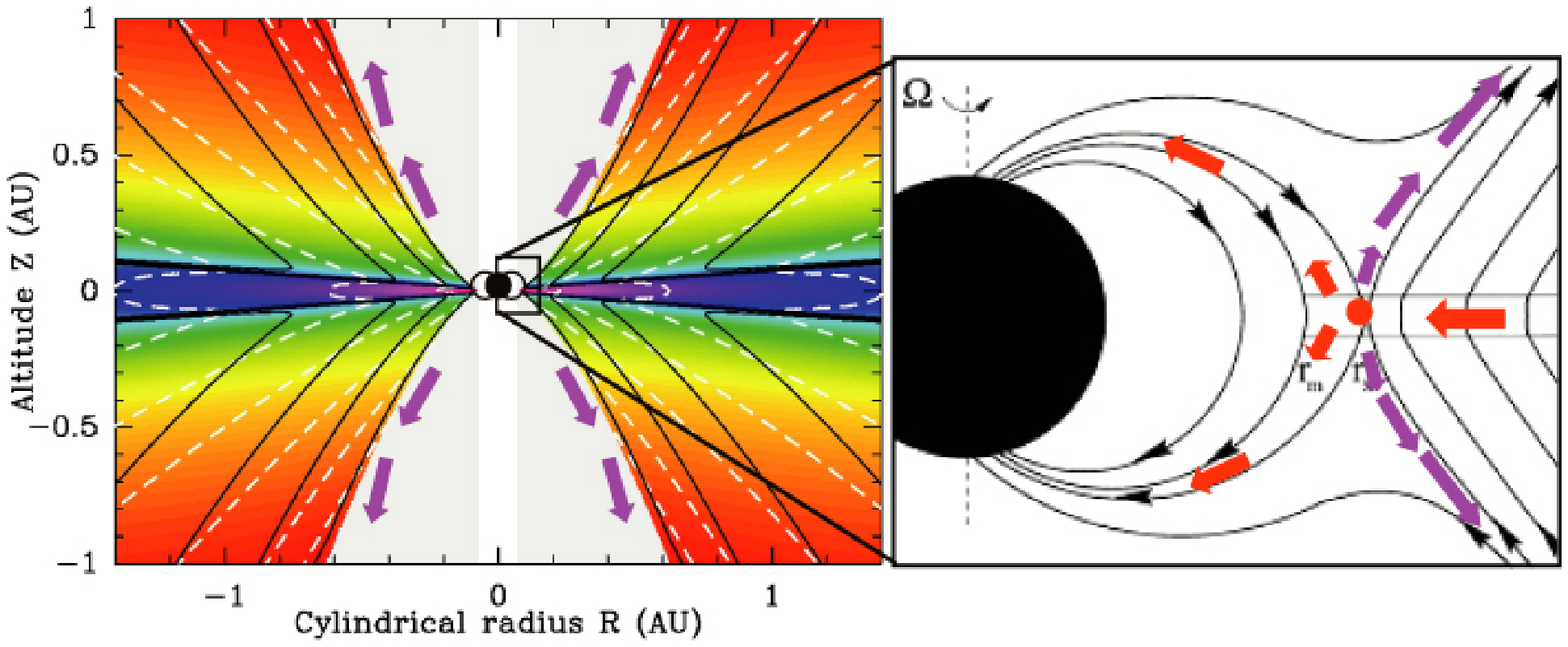}}
    \caption[]{Left: self-collimated jets from a keplerian accretion disc,
    with streamlines (black solid), contours of equal total velocity (white
    dashed) and density stratification in grayscale (Ferreira \cite{f97}). 
    Right: sketch of the magnetic configuration leading to "Reconnection
    X-winds" above the magnetic neutral line.  Arrows show the expected
    time-dependent plasma motion (Ferreira et al. \cite{f00}).}
\end{figure}

Still, observed jets are probably of no stellar origin.  But if jets are
indeed produced by the circumstellar disc, then the large scale
magnetic field in the disc has to match in some way the stellar
magnetospheric field.  We already know that such a matching has to occur in a
narrow region around $r_{m}$, where it is reasonable to assume a
stellar magnetic field of dipolar structure.  Now, if the stellar
magnetic moment is oriented in the same direction as the disc magnetic
field then one gets a situation where both magnetic fields cancel each
other (Figure~1).  This leads to the formation of a magnetic neutral
line\footnote{In this simplified picture, the magnetic moment is
aligned with the stellar rotation axis, leading to an axisymmetric
neutral line.  But this process (and its consequences) remains valid if the
dipole is misaligned.  In this case, the magnetic neutral line breaks
into two oppositely directed angular sectors with a magnetic neutral
segment.} at the axial distance $r_{X}$, defined as $B_{\star} (r_{X})
= B_{disc}(r_{X})$ (Ferreira et al. \cite{f00}).

This neutral line is an azimuthaly extended reconnection zone: closed
magnetospheric field lines are being "converted" into open field lines
linked to the star.  If the disc turbulence is indeed able to sustain
the amount of magnetic resistivity required, then such a picture could
well be a (time-averaged) representation of a star--disc interface. 
Note that this reconnection line is located in the disc equatorial
plane, so that it is not clear whether or not it could be observed as
some X-ray activity (ie.  flares related to the presence of a disc).

\subsection{Extraction of stellar angular momentum by a wind}

Where would this magnetic X-point be located in the disc~?  A
reasonable answer is provided if we demand that the model satisfies
(on a time-averave sense) all the following constraints: narrow
interface at $r_{m} \geq r_{co}$, magnetic braking by a stellar wind
and jet formation from the circumstellar disc.  The last constraint
tells us that the disc magnetic field must be close to equipartition,
namely $B_{disc} \propto \dot M_{acc}^{1/2}$.  The first constraint
obviously imposes $r_{X} > r_{m} \geq r_{co}$.  Using some crude
prescription for the stellar magnetic field, namely $B_{\star} \propto
r^{-n}$ ($n \geq 3$ free and mimicking a "compressed" magnetosphere),
then one gets the value of the stellar magnetic moment, which can then
be compared to observations.

The remarkable property of such a configuration is that magnetic braking is
automaticaly achieved.  Indeed, at radii greater than $r_{X}$, disc
plasma is unaware of the star and its dynamics is exactly the same as
in an accretion--ejection structure.  But around $r_{X}$ a sharp
transition occurs: matter on the disc equatorial plane crosses the
magnetic neutral line and is deflected vertically by the vertical
Lorentz force to form accretion curtains (at $r_{m}< r_{X}$ all disc
material has been deflected).  But matter which is already located at the
disc surface is pushed vertically away and loaded onto newly opened
field lines.  Now, those lines are rotating at the stellar angular
velocity which is greater than that of the plasma (as long as $r_{X} >
r_{co}$).  This means that this material is experiencing a
"magneto-centrifugal" acceleration from the star~!  Hence, the star
itself is indeed powering (with rotational energy and angular
momentum) plasma that was originally inside the disc.  Since plasma is being
loaded at $r_{X}$ and not at $R_{\star}$ as in conventional stellar
winds, we call these winds "Reconnection X-winds".

It is noteworthy that such ejection events at the star--disc interface
are most probably time-dependent.  They are indeed highly dependent on
the fact that $r_{X} > r_{co}$ and on the local magnetic topology, which
itself depends on both stellar dynamo and disc MHD turbulence.  So,
we expect that such a configuration is actually producing
accretion--ejection events in some intermittent way, involving
short (days: MHD turbulence) to quite long (years or more: stellar
dynamo) timescales.  Since every ejection event is channeled by the surrounding
disc-wind, it would observationaly appear as a "collimated" bullet.
 
Such a configuration is especially useful for the very early phase of
the star (embedded protostar, Class 0 and I).  Indeed, once optically
visible, the star rotates at only $10 \% $ of its break-up velocity. 
This means that stellar material, that originally came from a
molecular cloud, already lost a huge amount of specific angular
momentum.  To my knowledge, such an efficient angular momentum extraction has
never been obtained during the collapse itself.  It can be shown however that
Reconnection X-winds can very efficiently brake down a contracting
protostar (from break-up to $10 \% $ of it), on timescales compatible
with the duration of the embedded phase (Ferreira et al. \cite{f00}).

\subsection{Fossil fields and early stellar dynamo}

To summarize, we suppose that a non-negligible fraction of the magnetic flux
carried by the parent molecular cloud remains trapped during the
formation of the protostellar core.  The result is a protostar with a
magnetic moment parallel to the magnetic field threading the newly
formed surrounding accretion disc.  We expect that in the innermost
region of the disc the magnetic energy will be close to equipartition
with the disc thermal energy, initiating jets from the disc itself. 
We further suppose that a stellar dynamo begins to operate in the
rapidly rotating convective protostar.  In contrast with usual stellar
dynamo models, the protostar is in contact with an accretion disc
which itself carries a remnant magnetic flux.  Note that by contact,
we mean transfer of both mass and angular momentum: the behaviour of
such dynamos has not been studied yet.  We therefore speculate that
the particular boundary conditions provided by the magnetized disc
will favour growing dynamo modes with a magnetic moment aligned with
the disc field (on a quasi-stationary regime)\footnote{This behaviour
will obviously change once the disc has vanished.  In particular,
diffusion of any fossil field will then become possible.}.

Although this is a speculation, it is quite natural to envision protostar
evolution as being strongly influenced by both initial (presence of
strong fossil fields) and boundary (circumstellar accretion disc)
conditions.  Such a speculation leads to Reconnection X-winds, that may
solve the angular momentum problem in star formation.  If this picture
is viable, then dynamo theory would have to be revisited, for a
protostar could no longer be treated as a convective star in vacuum.

%%-----------------------------
%%      your bibliography
%%-----------------------------

\end{document}